\documentclass[twocolumn,prd,superscriptaddress,showpacs]{revtex4-2}
\usepackage{graphicx}
\usepackage{dcolumn}
\usepackage{amsmath}

\begin{document}

\title{Gravitational baryogenesis in Gauss-Bonnet braneworld cosmology}
\author{Ian B. Whittingham}
\affiliation{College of Science and Engineering,
James Cook University, Townsville, Queensland, Australia 4811}

\date{\today}

\begin{abstract}
One of the big unanswered questions of particle physics and cosmology is the origin of the small asymmetry in 
the abundance of matter and antimatter in the very early universe, described by the 
Baryon Asymmetry Factor $\eta \sim (8 - 9) \times 10^{-11}$. 

Gravitational baryogenesis postulates the baryon-antibaryon asymmetry is generated by a coupling between the baryon current 
and the gradient of the Ricci scalar $R$. As this coupling is zero in General Relativity (GR) for a radiation dominated flat early 
Universe, more generalized theories of gravity are required. 

The present study investigates the feasibility of gravitational baryogenesis in a Gauss-Bonnet (GB) modified Randall-Sundrum 
(RS) type II braneworld cosmology. The modified expansion rate of the Universe in this scenario evolves through three 
distinct expansion regimes, the GB regime at early times where the expansion rate $H \sim \rho^{1/3}$ 
falls more slowly than the standard expansion law $H\sim \rho^{1/2}$ of GR, then the RS type era with an enhanced 
expansion $H \sim \rho$, before finally reducing to the standard expansion law in the low energy limit. 
Gravitational baryogenesis depends on $\dot{R}$ and requires $\dot{R} <0$. This condition is satisfied for the GB and GR regimes.
However, $\dot{R}>0$ for the RS regime and successful baryogenesis requires either that the decoupling temperature occurs 
during the GB regime or that the Universe makes a direct transition from the GB regime to the GR regime.

The GB baryon asymmetry $\eta(T)$ is proportional to $\mu (1+w)(1-w)(1+1/\beta)$,  where $\mu$ is the energy scale $\mu \equiv 1/l$ associated with the bulk curvature scale $l$, $\beta \equiv 4 \alpha \mu^{2}$, where $\alpha$ is the GB coupling, and $w$ is 
the equation of state parameter $p=w \rho$. Studies of inflation in GB-braneworld cosmology place constraints on the allowed 
ranges of the parameters $\mu$ and $\beta$. Guided by these, the asymmetry has been calculated for ranges of $(\mu, \beta, w)$,
assuming a decoupling temperature $2 \times 10^{16}$ GeV and cutoff scale $M_{\mathrm{Pl}}$, and it is found that the 
observed values can be obtained for a value of $\beta$ within the range  $(10^{-2} - 10^{-3})$ for $\mu =10^{12}$ GeV, and 
within the range $(10^{-1}- 10^{-2})$ for $\mu =10^{13}$ GeV. All values of $\eta$ for $\mu=10^{14}$ are too high.
In contrast to GR where $\eta$ is zero for a radiation dominated Universe ($w=1/3$), it is non-zero in GB braneworld cosmology.
The parameter ranges for the decoupling temperatures required in the GB regime to produce the observed asymmetry have also been 
established.

\end{abstract}

\maketitle

\section{Introduction}

There is strong evidence that a matter-antimatter asymmetry exists in the Universe. Big Bang Nucleosynthesis (BBN), 
and measurements of the anisotropies of the Cosmic Microwave Background (CMB) combined with the large scale structure of the Universe, 
show that the number density of matter is of the order of $10^{-9}$ of that of the photon number density $n_{\gamma}$, 
but there is no antimatter in the present Universe other than that produced in cosmic ray collisions. 
This asymmetry is represented by the Baryon Asymmetry Factor
\begin{equation}
\label{IW0}
\eta \equiv \frac{n_{\mathrm{B}}-n_{\bar{\mathrm{B}}}}{s},
\end{equation}
where $n_{\mathrm{B}}(n_{\bar{\mathrm{B}}})$ is the number density of baryons (antibaryons), and $s$ is the entropy density of the relativistic radiation bath. 
Current observations~\cite{Cooke2014,Ade2016} give
\begin{equation}
\label{IW1}
\eta  = \frac{1}{7.04} \frac{n_{\mathrm{B}}-n_{\bar{\mathrm{B}}}}{n_{\gamma}}
=\left\{
			\begin{array}{ll}
			(8.2-9.4) \times 10^{-11} & (\mathrm{BBN}) \\
			(8.65 \pm 0.09) \times 10^{-11} & (\mathrm{CMB})
			\end{array} \right.
\end{equation}

Conventionally, this asymmetry is generated dynamically from an initial baryon-symmetric phase in the primordial Universe. 
Sakharov~\cite{Sak1967} established the necessary conditions for such a generation of baryon asymmetry: 
(i) Baryon-number non-conserving interactions, (ii) $C$ and $CP$ violation and (iii) departure from thermal equilibrium.

One popular mechanism for departure from thermal equilibrium is the electroweak phase transition which occurs during the expansion of the Universe at $T \sim 100$ GeV. This has been studied both within the Standard Model~\cite{Vis2025} and, particularly, its extension to include an additional Higgs doublet, possibly arising from supersymmetry (see, eg, \cite{ACW1993}).

An alternative mechanism for generating a baryon asymmetry is Gravitational Baryogenesis~\cite{DKKMS2004}, in which
dynamical breaking of $CPT$ (and $CP$) occurs during the expansion of the universe and thermal equilibrium is preserved. 
It postulates a gravitational interaction between the baryon number current $j^{\mu}_{\mathrm{B}}$ and the derivative 
$\partial_{\mu} R$ of the Ricci scalar $R$, with the action
\begin{eqnarray}
\label{IW2}
\mathcal{S}_{\mathrm{B}} & = & \frac{1}{M_{\ast}^{2}} \int d^{4}x \sqrt{-g} (\partial _{\mu}R) j^{\mu}_{\mathrm{B}} \\
                          & \equiv & \int d^{4}x \sqrt{-g} \mathcal{L}_{\mathrm{B}},
\end{eqnarray}
where $g=\det (g_{\mu \nu})$ and  $g_{\mu \nu}$ is the metric tensor.
This interaction biases the dynamics in the expanding Universe, causing an asymmetry between 
particles and antiparticles. It requires the $B$-violating processes to be in thermal equilibrium and modifies the
thermal equilibrium distributions in a similar fashion to a chemical potential. 
$B$-violation decouples below a temperature $T_{\mathrm{D}}$. 

An interaction of the form $\mathcal{L}_{\mathrm{GB}}$ is expected in the low-energy effective field theory of quantum gravity 
if the cutoff scale $M_{\ast}$ of the effective theory is of order of the reduced Planck scale~\cite{DKKMS2004}
\begin{equation}
\label{IW3}
M_{\mathrm{Pl}} = (\hbar c/(8 \pi G_{\mathrm{N}}))^{1/2} = 2.435 \times 10^{18} \, \mathrm{GeV}/c^{2}.
\end{equation}

For an expanding spatially uniform homogeneous universe the interaction reduces to
\begin{equation}
\label{IW4}
\mathcal{L}_{\mathrm{B}} = -\frac{1}{M_{\ast}^{2}} \dot{R} \, j^{0}_{\mathrm{B}}.
\end{equation}
where
\begin{equation}
\label{IW8}
j^{0}_{\mathrm{B}} = n_{\mathrm{B}} - n_{\bar{\mathrm{B}}}.
\end{equation}
This interaction (\ref{IW4}) has the form
\begin{equation}
\label{IW37}
\mathcal{L}_{\mathrm{B}} = \mu \,\delta n_{\mathrm{b}},
\end{equation}
where
\begin{equation}
\label{IW39}
\mu \equiv - \frac{1}{M_{\ast}^{2}} \dot{R}
\end{equation}
is the effective chemical potential for the system and, for a system of ultrarelativistic fermions in thermal equilibrium 
at temperature $T$, their Fermi-Dirac distributions give
\begin{equation}
\label{IW38}
\delta n_{\mathrm{b}} \equiv n_{\mathrm{B}} - n_{\bar{\mathrm{B}}}
= \frac{g_{\mathrm{B}}}{6 \pi^{2}} \left( \frac{k_{\mathrm{B}}T}{\hbar c} \right)^{3} \,
		\left[ \frac{\mu}{k_{\mathrm{B}}T}\,\pi^{2} + \left(\frac{\mu}{k_{\mathrm{B}}T}\right)^{3}\right],
\end{equation}
where $g_{\mathrm{B}}=2$ is their number of spin states.

The entropy density of the early Universe for all particles in thermal equilibrium at temperature $T$ is
\begin{equation}
\label{IW40}
s = \frac{k_{\mathrm{B}}^{4}}{(\hbar c)^{3}}\, \frac{2 \pi^{2}}{45} g_{\ast}(T) T^{3},
\end{equation}
where $g_{\ast}(T)$ is the number of degrees of freedom of ultrarelativistic particles present, and is 
equal to 106.75 for $k_{\mathrm{B}}T \gg m_{t}c^{2}$. The Baryon Asymmetry Factor to order $\mu$ is therefore
\begin{equation}
\label{IW41}
\eta = \frac{\delta n_{\mathrm{b}}}{s} = -\left( \frac{15 g_{\mathrm{B}}}{4 \pi^{2} k_{\mathrm{B}}g_{\ast}(T)} \right)
\;\frac{\dot{R}}{M_{\ast}^{2}T}.
\end{equation}
Once $T$ drops below the decoupling temperature $T_{\mathrm{D}}$, this ratio is frozen in at the value $\eta|_{T=T_{\mathrm{D}}}$.

In General Relativity (GR), Einstein's equations  
\begin{equation}
\label{IW43}
R_{\mu \nu} -\frac{1}{2} g_{\mu \nu} R = \kappa^{2} T_{\mu \nu},
\end{equation}
where $R_{\mu \nu}$ is the Ricci curvature tensor, $T_{\mu \nu}$ is the matter energy-momentum tensor, 
and $\kappa^{2}=8\pi G_{\mathrm{N}} = 1/M_{\mathrm{Pl}}^{2}$, give, for a flat Friedmann-Robertson-Walker (FRW) Universe,
$R=-\kappa^{2}T =-\kappa^{2} \rho (1-3w)$, where $w$ is the equation of state parameter $p=w \rho$, 
$R = R^{\mu}_{\mu}$ is the Ricci scalar, and $T=T^{\mu}_{\mu}$.  Using the energy conservation equation
\begin{equation}
\label{IW50a}
\frac{d \rho}{dt} + 3 H (\rho+p)=0,
\end{equation}
gives
\begin{equation}
\label{IW300}
\dot{R} =-\sqrt{3} \kappa^{3} (1-3w) (1+w) \rho^{3/2},
\end{equation}
and the Baryon asymmetry factor becomes
\begin{eqnarray}
\label{IW303}
\frac{\delta n_{\mathrm{b}}}{s} & = & \sqrt{3} 
\left( \frac{15 g_{\mathrm{B}}(1-3w)(1+w)}{4 \pi^{2}g_{\ast}(T)M_{\ast}^{2} M_{\mathrm{Pl}}^{3} T} \right)\rho^{3/2}.
\end{eqnarray}
Hence, in GR, $\dot{R}$ is zero for a radiation-dominated early Universe where $w=1/3$, so more generalized 
theories of gravity need to be considered in order for gravitational baryogenesis to be possible.

Gravitational baryogenesis has been investigated for a wide range of gravitational models. 
The early calculation by Davoudiasl et al \cite{DKKMS2004} is based upon GR but  considers three scenarios to realize $w \neq 1/3$. 
The first is a radiation-dominated era following inflation when $w \approx 1/3$. 
In the limit of exact conformal invariance, $T^{\mu}_{\mu}=0$ and $w=1/3$. However, interactions among massless particles lead to 
running coupling constants and hence the trace anomaly $T^{\mu}_{\mu}\neq 0$.
Typical gauge groups and matter content at very high energies give $1-3w \sim 10^{-2} - 10^{-1}$.
The second scenario is a matter-dominated era $w=0$ such as reheating via a scalar field $\phi$
as it oscillates about the minimum of a quadratic potential $V(\phi) \propto \phi^{2}$~\cite{Turner1983}. 
The third scenario is to generate the asymmetry while a non-thermal component with $w > 1/3$ dominates the 
Universe. A general $w$ can be realized, for example, by a coherent oscillation of $\phi$ about the minimum of 
a potential $V(\phi) \propto \phi^{n>2}$~\cite{Turner1983}.

Many investigations are based upon $f(R)$ gravity~\cite{NOO2017}, in which the gravitational  sector of the 
Einstein-Hilbert action of GR
\begin{equation}
\label{IW46}
\mathcal S _{\mathrm{EH}} =  \int d^{4}x \sqrt{-g} [\frac{1}{2 \kappa^{2}} R +\mathcal L_{m}(g_{\mu \nu},\psi_{m})],
\end{equation}
is generalized to
\begin{equation}
\label{IW46aa}
\mathcal S  = \int d^{4}x \sqrt{-g} [\frac{1}{2 \kappa^{2}}f(R) +\mathcal L_{m}(g_{\mu \nu},\psi_{m})].
\end{equation}
Here, $\mathcal L_{m}$ is Lagrangian density for the matter fields $\psi_{m}(x)$.

Lambiase and Scarpetta\cite{LS2006} consider baryogenesis in a $f(R)$ gravity with the simple power law~\cite{KKKC2015} 
\begin{equation}
\label{IW160}
f(R) = \left(\frac{R}{A}\right)^{n}.
\end{equation}
This has been generalized by Ramos and P\'{a}ramos~\cite{RP2017} to nonminimally coupled $f(R)$ gravity where 
\begin{equation}
\label{IW160a}
\mathcal S  = \int d^{4}x \sqrt{-g} [\frac{1}{2 \kappa^{2}}f_{1}(R) +f_{2}(R)\mathcal L_{m}(g_{\mu \nu},\psi_{m})].
\end{equation}
with the form
\begin{equation}
\label{IW161}
f_{1}(R)=R\left(\frac{|R|}{M_{1}^{2}}\right)^{m}, \quad f_{2}(R)=R\left(\frac{|R|}{M_{2}^{2}}\right)^{n}.
\end{equation}
Several studies~\cite{Pizza2015,OOS2023,ABS2010,Whitt2026} have considered the Starobinsky model~\cite{Star1980}
\begin{equation}
\label{IW333}
f(R) = R + \frac{R^{2}}{M^{2}}
\end{equation}
or its modifications.  The model is strongly motivated by theoretical considerations of corrections to Einstein's GR 
arising from quantum corrections and new physics. Measurements of the primordial scalar spectrum of the CMB~\cite{Planck2018a} 
support the Starobinsky model for $M/M_{\mathrm{P}} \simeq 1.3 \times 10^{-5}$.

Gravitational baryogenesis has been studied in other modified gravities.  
Odintsov and Oikonomou consider Gauss-Bonnet baryogenesis~\cite{OO2016} based on the coupling 
$j^{\mu}_{\mathrm{B}} \partial_{\mu}G$, where
\begin{equation}
\label{IW47} 
G =R^{2}-4R_{\mu\nu}R^{\mu\nu}+R_{\mu\nu\rho\sigma}R^{\mu\nu\rho\sigma}
\end{equation} 
is the Gauss-Bonnet curvature invariant constructed from the Riemann tensor $R_{\mu\nu\rho\sigma}$. 
Odintsov and Oikonomou also consider~\cite{OO2016a} the effects of Loop Quantum Cosmology on the interaction  
$j^{\mu}_{\mathrm{B}} \partial_{\mu}R$, for an Einstein-Hilbert Universe with a flat FRW metric. These effects produce an 
additional term in the GR expression (\ref{IW300}) for $\dot{R}$ proportional to $(1+3w)\rho^{2}/\rho_{c}$, 
where the parameter $\rho_{c}$ represents the quantum effects of the theory.  
Oikonomou and Saridakis~\cite{OS2016} have studied 
baryogenesis in $f(\mathcal{T})$ gravity, where $R$ is replaced by the torsion scalar $\mathcal{T}$.

These models assume the baryon asymmetry arises from gravitational corrections to the Einstein-Hilbert action.
As alternatives, Pereira et al have studied baryogenesis generated by higher-order matter contributions incorporated in a 
$f(R,T^{2})$ gravity~\cite{PLM2025}, and by gravitational scalars $\phi_{i}$ in a scalar-tensor representation of 
modified gravity~\cite{Pereira2026}. 

Gravitational baryogenesis has been considered by Sahoo \textit{et al} for several extended models: $f(R,T)$ ~\cite{SB2020},
$f(Q,T)$~\cite{BS2020}, where $Q$ is the nonmetricity, $f(R,\mathcal{L}_{m})$~\cite{JBS2023}, 
$f(\mathcal{T},T)$~\cite{MMS2023}, and $f(Q)$~\cite{MBS2024}. 

The present study investigates the feasibility of gravitational baryogenesis in a Gauss-Bonnet (GB) modified Randall-Sundrum 
type II braneworld cosmology. This scenario has been considered by Bento \textit{et al}~\cite{BFS2005} but only for the
restricted case of a radiation dominated Universe. 

The structure of the paper is as follows. The basic equations of Gauss-Bonnet braneworlds are summarized in Sec. 2 and the three expansion 
regimes of the Universe discussed, expressions for gravitational baryogenesis are obtained in Sec. 3, and the calculations of 
gravitational baryogenesis for the GB expansion regime are given in Sec. 4. Sec. 5 contains a summary and conclusions 
for the investigation.

\section{Gauss-Bonnet Braneworlds}

A braneworld cosmological scenario assumes the observable universe is a 3(+1) dimensional surface (the 'brane') 
embedded in a five dimensional bulk spacetime. 
Standard Model particles are confined to the surface of the brane whilst gravity propagates in the higher dimensional 
bulk~\cite{Langlois2003,Maartens:2010ar}. This class of models is motivated by (super)string theory and M-theory which require additional spacetime dimensions for internal consistency.

In the widely studied Randall-Sundrum type II (RSII) model~\cite{Randall:1999vf}, the 5-dimensional action is Einstein-Hilbert and
GR is recovered on the surface of a 3(+1) Minkowski brane located at the ultraviolet boundary of a 5-dimensional 
anti-de Sitter  (AdS$_{5}$) bulk. 
The warped geometry of the bulk spacetime ensures the fifth dimension is only accessible in the ultraviolet regime and that 
$\Lambda$CDM is reproduced in the low energy limit. 

In this article we consider an extension of the RSII model which incorporates the Gauss-Bonnet (GB) higher order curvature 
term  $\mathcal{G}$~\cite{Gross198741} in the bulk action integral, thus modifying the braneworld dynamics at high energies.
Inclusion of higher order curvature terms generally leads to fourth order equations of motion. However,
in five dimensions, the GB combination of invariants $\mathcal{G}$ constructed from the 5-dimensional Riemann tensor 
$\mathcal{R}_{abcd}$ is of particular significance since it is the unique combination that leads to second order gravitational field equations in the bulk metric which are symmetric, divergenceless and ghost free~\cite{Clifton:2011jh}.
The inclusion of a GB term affects early universe inflation and modifies the power spectrum of both scalar and tensor 
primordial perturbations and the consistency relation between 
them~\cite{DLMS2004,Tsujikawa2004a,Tsujikawa2004b,Calcagni2013,Neupane2014,OO2015,CPS2018}.

The gravitational action for the GB braneworld model, taken over the five dimensional bulk spacetime $\mathcal{M}$, is 
\begin{equation}
\label{IW48}
S_\mathcal{M} = \frac{1}{2\kappa_5^2}\int_\mathcal{M}{d^5x\,\sqrt{-g_{5}}\left[\mathcal{R} - 2\Lambda_5 + \alpha\mathcal{G}\right]},
\end{equation}
where $g_{5}$ is the determinant of the bulk metric $g_{ab}$, $\mathcal{R}$ is the five dimensional Ricci scalar and $\Lambda_5(<0)$ 
is the bulk cosmological constant. The GB contribution is parametrized through the coupling $\alpha$ which, if GB gravity is
regarded as the lowest order stringy correction to the 5-dimensional Einstein-Hilbert action, satisfies 
$\alpha |\mathcal{R}|^{2} \ll |\mathcal{R}|$,
so that $\alpha \ll l^{2}$, where $l$ is the bulk curvature scale $|R| \sim l^{-2}$, and is therefore positive.

The matter fields, which are localized on the brane surface $\partial\mathcal{M}$, are included via
\begin{equation}
\label{IW48a}
S_m =  - \int_{\partial\mathcal{M}}{d^4x\,\sqrt{-g_{4}}\,[\mathcal{L}_{m} + \sigma]},
\end{equation}
where $g_{4}$ is the determinant of the induced metric $g_{\mu \nu}$ on the brane surface, $\mathcal{L}_m$ is the matter field 
Lagrangian and $\sigma(>0)$ is the brane tension. Varying the total action $S_{tot} = S_\mathcal{M} + S_{m}$ (+ boundary terms) 
with respect to the metric field and solving the resulting field equations yields, for a flat FRW brane, 
the modified Friedmann equation~\cite{CD2002,Davis2003}
\begin{equation}
\label{IW50}
\kappa_5^2\left(\rho + \sigma\right) = 2\sqrt{\mu^{2} + H^2}\left(3 - 4 \alpha \mu^{2} + 8 \alpha H^2\right),
\end{equation}
where $\rho$ is the energy density of matter fields on the brane. For a vacuum bulk the brane energy-momentum tensor is 
conserved, $\nabla^{\nu}T_{\mu \nu}=0$, and, for a perfect fluid matter source, the conservation equation (\ref{IW50a}). 

For an AdS$_{5}$ bulk,
\begin{eqnarray}
\label{IW49}
\Lambda_{5} & = & -\frac{6}{l^{2}} + \frac{12 \alpha}{l^{4}}, 
\nonumber  \\
\frac{1}{l^{2}} & \equiv & \mu^{2} = 
\frac{1}{4 \alpha}\left[1- \sqrt{1+ 4\alpha\Lambda_5/3} \right],
\end{eqnarray}
providing the upper limit on the GB coupling $\alpha < l^{2}/4$, which ensures $\Lambda_{5}<0$.

\cite{Neupane2014} gives the following constraints on GB Braneworld inflation: 
(i) Inflationary constraints $(-\alpha \Lambda_{5}/3) < O(10^{-3})$, (ii) Stability of classical solutions under perturbation 
$(-\alpha \Lambda_{5}/3) < O(1/10)$ , and (iii) BICEP2~\cite{BICEP2} and Planck $(-\alpha \Lambda_{5}/3) < (3-300) \times 10^{-5}$.

For $p=w \rho$, the GR conservation equation (\ref{IW50a}) gives
\begin{equation}
\label{IW50b}
H = \frac{d \ln a}{dt} = \frac{d \ln a}{d \rho} \frac{d \rho}{dt} = \frac{d \ln a}{d \rho} [-3H (1+w)\rho],
\end{equation}
with the solution
\begin{equation}
\label{IW50c}
\rho(a) = \rho_{0} a^{-3(1+w)}.
\end{equation}

As shown in~\cite{LN2003}, it is possible to solve (\ref{IW50}) to get an explicit expression for the Hubble factor $H$;
\begin{equation}
\label{IW51}
H^2 = \frac{1}{4\alpha}\left[\left(1 - 4\alpha \mu^{2}\right)\cosh{\left(\frac{2\chi}{3}\right)} - 1\right],
\end{equation}
where $\chi$ is related to the energy density $\rho$ via
\begin{equation}
\label{IW52}
\rho + \sigma = m_\alpha^4\sinh{\chi},
\end{equation}
and
\begin{equation}
\label{IW53}
m_{\alpha} = \left[\frac{2(1-4 \alpha \mu^{2})^{3}}{\alpha \kappa_{5}^{4}}\right]^{1/8}.
\end{equation}
To see this, substitution of (\ref{IW51}) into (\ref{IW50}) and using the identity 
\begin{equation}
\label{IW53a}
\sinh \chi = 4 \sinh^{3} \left(\frac{\chi}{3} \right)+ 3 \sinh \left(\frac{\chi}{3}\right)
\end{equation}
gives
\begin{eqnarray}
\label{IW54}
\kappa_{5}^{2}(\rho + \sigma) & = & \left[\frac{(1-4\alpha \mu^{2})^{3}}{\alpha}\right]^{1/2}
\left[\cosh\left(\frac{2 \chi}{3}\right)-1\right]^{1/2}
\nonumber  \\
& & \times \left[2\cosh\left(\frac{2 \chi}{3}\right)+1\right]
\nonumber  \\
& = &  m_{\alpha}^{4}
 \left[4 \sinh^{3}\left(\frac{\chi}{3}\right) + 3 \sinh \left(\frac{\chi}{3}\right)\right]
\nonumber \\
& = &  m_{\alpha}^{4} \sinh \chi .
\end{eqnarray}

Imposing the condition $H=0$ when $\rho=0$ requires, from (\ref{IW51}),
\begin{equation}
\label{IW55}
\cosh{\left(\frac{2\chi}{3}\right)} = \frac{1}{1-4\alpha \mu^{2}}.
\end{equation}
Using
\begin{equation}
\label{IW56}
\sinh^{2}\left(\frac{\chi}{3}\right) = \left[\frac{2 \alpha \mu^{2}}{1-4 \alpha \mu^{2}}\right].
\end{equation}
then (\ref{IW53a}) gives
\begin{equation}
\label{IW57}
\sinh \chi  =  \left[\frac{2 \alpha \mu^{2}}{(1-4 \alpha \mu^{2})^{3}}\right]^{1/2}[3-4\alpha \mu^{2}].
\end{equation}
From (\ref{IW52}), this corresponds to the brane tension
\begin{equation}
\label{IW58}
\kappa_{5}^{2}\sigma = 2 \mu (3-4 \alpha \mu^{2}).
\end{equation}

The modified Friedmann equation (\ref{IW50}) clearly predicts non-standard behaviour for the 
expansion of the universe, as characterized by the GB regime $\rho \gg m_{\alpha}^{4}$ and the RS regime $ m_{\alpha}^{4} \gg \rho$.
Note that, since the GB term is a high energy correction to the regular RS action, we expect $m_{\alpha}^{4} \gg \sigma$. 

For $\rho \gg m_{\alpha}^{4}$, (\ref{IW52}) gives
\begin{equation}
\label{IW59}
\sinh \chi = \frac{\rho + \sigma}{m_{\alpha}^{4}} \approx \frac{\rho}{m_{\alpha}^{4}} \gg 1.
\end{equation}
Hence
\begin{equation}
\label{IW60}
\sinh \chi \approx \frac{1}{2} e^{\chi}, \quad \cosh \left(\frac{2 \chi}{3}\right) \approx \frac{1}{2} e^{2 \chi/3}
\end{equation}
and, from (\ref{IW51}), the GB expansion
\begin{equation}
\label{IW61}
H^{2} = \left[\frac{\kappa_{5}^{2}}{16 \alpha}\right]^{2/3}  \rho^{2/3}.
\end{equation}

For $m_{\alpha}^{4} \gg \rho$, (\ref{IW52}) gives
\begin{equation}
\label{IW62}
\sinh \chi = \frac{\rho + \sigma}{m_{\alpha}^{4}} \ll 1
\end{equation}
and therefore
\begin{eqnarray}
\label{IW63}
\sinh \chi \approx \chi  & = & \frac{\rho + \sigma}{m_{\alpha}^{4}}, \nonumber  \\
\cosh \left(\frac{2 \chi}{3}\right) & \approx & 1+ \frac{2 \chi^{2}}{9} = 
1+\frac{2}{9} \left(\frac{\rho + \sigma}{m_{\alpha}^{4}}\right)^{2}.
\end{eqnarray}
From (\ref{IW51}), the Hubble factor is
\begin{equation}
\label{IW64}
H^{2}=\frac{\kappa_{5}^{4}}{36  (1-4 \alpha \mu^{2})^{2}} (\sigma + \rho)^{2} - \mu^{2},
\end{equation}
which, applying the condition (\ref{IW58}), becomes
\begin{equation}
\label{IW65}
H^{2}= \frac{1}{3}\left[\frac{\mu \kappa_{5}^{2}}{1+4 \alpha \mu^{2}}\right] \rho c_{1} \left(1+\frac{\rho}{2 \sigma}\right) +c_{2},
\end{equation}
where
\begin{equation}
\label{IW66}
c_{1} \equiv \frac{(1+4 \alpha \mu^{2})(3-4 \alpha \mu^{2})}{3(1-4 \alpha \mu^{2})^{2}} \approx 1
\end{equation} 
and 
\begin{equation}
\label{IW67}
c_{2}\equiv \frac{1}{9} \left(\frac{3-4\alpha \mu^{2}}{1-4 \alpha \mu^{2}}\right)^{2} \mu^{2} -\mu^{2} \approx 0
\end{equation}
since $\alpha \mu^{2} \ll 1$. Hence
\begin{equation}
\label{IW68}
H^{2}= \frac{1}{3}\left[\frac{\mu \kappa_{5}^{2}}{1+4 \alpha \mu^{2}}\right] \rho \left(1+\frac{\rho}{2 \sigma}\right),
\end{equation}
which, in the low energy regime  $\rho \ll \sigma$, gives the GR Friedmann equation
\begin{equation}
\label{IW69}
H^{2}= \frac{\kappa_{4}^{2}}{3} \rho
\end{equation}
if we identify
\begin{equation}
\label{IW70}
\kappa_4^2 =  \frac{\mu}{1 + 4\alpha \mu^{2}}\kappa_5^2.
\end{equation}
Note that the condition (\ref{IW58}) has also made the four dimensional cosmological constant $\Lambda_{4}$ vanish.

It is convenient to introduce
\begin{equation}
\label{IW71}
\beta \equiv 4 \alpha \mu^{2} =1-\sqrt{1+\frac{4 \alpha \Lambda_{5}}{3}} \ll 1, \quad m_{\sigma}^{4} \equiv \sigma.
\end{equation}
The two mass scales $m_\alpha$ and $m_\sigma$, which correspond to the GB correction and the brane tension respectively, 
can be written in terms of the two remaining free parameters $\mu$ and $\beta$ as
\begin{equation}
\label{IW72}
m_\alpha^4 = 2\,\frac{\mu^2}{\kappa_4^2}\sqrt{\frac{2(1 - \beta)^3}{\beta(1 + \beta)^2}},\quad 
m_\sigma^4 = 2\,\frac{\mu^2}{\kappa_4^2}\left(\frac{3-\beta}{1 + \beta}\right).
\end{equation} 
The ratio of the two mass scales
\begin{equation}
\label{IW73}
\mathcal{R}_{m} \equiv \frac{m_\alpha}{m_\sigma} = \left[\frac{2(1 - \beta)^3}{\beta(3 - \beta)^2}\right]^{1/8}
\end{equation}
depends only on $\beta$ and has the properties~\cite{MW2014}: (i) $\mathcal{R}_{m}=1$ for
\begin{equation}
\label{IW73a}
3 \beta^{3} - 12 \beta^{2} +15 \beta -2 =0,
\end{equation}
that is $\beta =0.1509$, and (ii) $\mathcal{R}_m \geq 1$ for $0 < \beta \leq 0.1509$.

The results (\ref{IW61}) and (\ref{IW68}) indicate that modified expansion rate in the generalized GB scenario evolves through 
three distinct expansion regimes, characterized by the mass scales $m_\alpha$ and $m_\sigma$~\cite{DLMS2004,OO2009}:
\begin{enumerate}
\item
The GB regime: $\rho \gg m_\alpha^4$
\begin{equation}
\label{IW74}
H^2 \simeq \left(\frac{1 + \beta}{4\beta}\mu\kappa_4^2\rho\right)^{2/3},
\end{equation}
\item
The RS regime: $m_\alpha^4\gg \rho \gg m_\sigma^4$
\begin{equation}
\label{IW75}
H^2 \simeq \frac{\kappa_4^2}{6m_\sigma^4}\rho^2,
\end{equation}
\item
The standard regime: $m_\sigma^4 \gg \rho$
\begin{equation}
\label{IW76}
H^2 \simeq \frac{\kappa_4^2}{3}\rho.
\end{equation}
\end{enumerate}
At early times, during the GB regime, the expansion rate of the universe $H \sim \rho^{1/3}$ falls more slowly than the standard expansion law $H\sim \rho^{1/2}$. Later, the universe evolves into a RS type era with an enhanced expansion $H \sim \rho$, before finally reducing to the standard expansion law in the low energy limit. 
The duration of the RS regime is determined by the magnitude of $\mathcal{R}_m$: 
when $\mathcal{R}_m$ is small, the RS era is short and the expansion rate passes quickly from the GB era to the standard era; 
when $\mathcal{R}_m$ is large, the duration of the RS era is extended. From (\ref{IW73}), these two cases correspond 
to $\beta\lesssim 0.1509$ and $\beta \rightarrow 0$, respectively. Since the GB term is a high energy correction to 
the RS action, then $m_\alpha \gg m_\sigma$, corresponding to $\beta \ll 1$. 

The modified Friedmann equation (\ref{IW68}) for the RS-GR regimes
\begin{equation}
\label{IW88}
H^{2}= \frac{\kappa_{4}^{2}}{3} \rho \left(1+\frac{\rho}{2 \sigma}\right),
\end{equation}
has the same form as that for Loop Quantum Cosmology~\cite{AS2011,WE2013,CW2015}
\begin{equation}
\label{IW89}
H^{2}= \frac{\kappa_{4}^{2}}{3} \rho \left(1-\frac{\rho}{\rho_{c}}\right),
\end{equation}
which describes a non-singular bounce cosmology, where $\rho_{c}$ is the maximum energy density. 
Since the $\rho^{2}$ corrections are of the opposite sign in braneworld cosmology to those in Loop Quantum Cosmology, 
bounce scenarios can only occur if the brane tension is allowed to be negative. 

These modified Friedmann equations, combined with the conservation equation (\ref{IW50a}), give
\begin{equation}
\label{IW90}
R= 12 H^{2} + 6 \dot{H} = \kappa_{4}^{2} \rho [(1-3w) + s_{\rho} (1+3w) \rho],
\end{equation}
where $s_{\rho}=2/\rho_{c}$ for Loop Quantum Cosmology and $s_{\rho}=-1/\sigma$ for braneworld cosmology. Since $\dot{\rho}<0$, 
then $\dot{R} <0$ for Loop Quantum Cosmology, but, for the RS regime of braneworld cosmology, $\dot{R} >0$ since 
$\rho \gg \sigma$ and the $(1+3w)$ term dominates.

\section{Gravitational baryogenesis}

Gravitational baryogenesis depends on $\dot{R}$ and requires $\dot{R} <0$. 
It is assumed that the interaction $\mathcal{L}_{\mathrm{B}}$ does not affect the cosmological background dynamics, the 
so called \emph{spectator approximation}~\cite{PFM2026}.

The modified Friedmann equations for the three regimes can be written in the generic form
\begin{equation}
\label{IW74aa}
H^{2} = \lambda_{\gamma} \rho^{\gamma}
\end{equation}
where
\begin{enumerate}
\item
The GB regime: $\rho \gg m_\alpha^4$
\begin{equation}
\label{IW74a}
\lambda_{\gamma} = \left(\frac{1 + \beta}{4\beta}\mu\kappa_4^2\right)^{2/3}, \quad \gamma=\frac{2}{3},
\end{equation}
\item
The RS regime: $m_\alpha^4\gg \rho \gg m_\sigma^4$
\begin{equation}
\label{IW75a}
\lambda_{\gamma} = \frac{\kappa_4^2}{6m_\sigma^4}, \quad \gamma=2,
\end{equation}
\item
The standard regime: $m_\sigma^4 \gg \rho$
\begin{equation}
\label{IW76a}
\lambda_{\gamma} = \frac{\kappa_4^2}{3}, \quad \gamma =1.
\end{equation}
\end{enumerate}

Combining (\ref{IW74aa}) with the conservation of energy-momentum result (\ref{IW50c}) gives
\begin{equation}
\label{IW77}
H^2 = \lambda_{\gamma} \rho_{0}^{\gamma}  a^{-\nu \gamma},
\end{equation}
where $\nu \equiv 3(1+w)$, that is
\begin{equation}
\label{IW78}
\frac{da}{dt} = \sqrt{\lambda_{\gamma} \rho_{0}^{\gamma}} a^{1-\nu \gamma/2}.
\end{equation}
Integrating $dt/da$ gives
\begin{equation}
\label{IW79}
a(t) = \left[\frac{\nu \gamma \sqrt{\lambda_{\gamma} \rho_{0}^{\gamma}}}{2} t \right]^{2/\nu \gamma},
\end{equation}
where we have taken $t_{0}=0$ and $a(t_{0})=0$.

The Hubble factor is then
\begin{equation}
\label{IW80}
H(t)= \frac{2}{\nu \gamma t},
\end{equation}
giving
\begin{equation}
\label{IW81}
R(t) = 12 H^{2} + 6 \dot{H} = \frac{12}{\left[\nu \gamma t \right]^{2}} (4-\nu \gamma)
\end{equation}
and
\begin{equation}
\label{IW82}
\dot{R}(t) = - \frac{24}{\left[\nu \gamma \right]^{2}} (4-\nu \gamma) \frac{1}{t^{3}}.
\end{equation}

Note that, from (\ref{IW82}) and (\ref{IW80}),
\begin{equation}
\label{IW82a}
\dot{R} = -3 \nu \gamma (4- \nu \gamma) H^{3}.
\end{equation}
For the case of a radiation dominated Universe $(\nu =4)$, this gives
\begin{equation}
\label{IW82b}
\dot{R} = -48 \gamma (1- \gamma) H^{3},
\end{equation}
in agreement with the result found by \cite{BFS2005}.

Explicitly, the values of $\dot{R}$ for the three regimes are
\begin{enumerate}
\item
The GB regime: 
\begin{equation}
\label{IW82a}
\dot{R} = - \frac{12 (1-w)}{(1+w)^{2} t^{3}}.
\end{equation}
\item
The RS regime: 
\begin{equation}
\label{IW82b}
\dot{R} = \frac{4(1+3w)}{3(1+w)^{2} t^{3}}.
\end{equation}
\item
The GR regime: 
\begin{equation}
\label{IW82c}
\dot{R} = - \frac{8 (1-3w)}{3(1+w)^{2} t^{3}}.
\end{equation}

\end{enumerate}

Gravitational baryogenesis requires the Universe to be in thermal equilibrium. 
If the $B$ number-violating interactions are generated by an operator $\mathcal{O}_{\mathrm{B}}$ of mass dimension 
$D=4+n$~\cite{DKKMS2004}, the rate of such interactions is given by
$\Gamma_{\mathrm{B}} =T^{2n+1}/M^{2n}_{\mathrm{B}}$, where $M_{\mathrm{B}}$ is the mass scale associated 
with $\mathcal{O}_{\mathrm{B}}$, and should be $\geq 10^{14}$ GeV to satisfy requirements from GUT or Planck-suppressed
interactions.
Decoupling of $B$-violating processes occurs when $\Gamma_{\mathrm{B}} > H$, requiring 
$M_{\mathrm{B}} \lesssim (T^{2n+1}/H)^{1/2n}$. For a dimemsion-six operator, this requires  
$M_{\mathrm{B}} \lesssim (T^{5}/H)^{1/4}$. 
Assuming this thermalization then
\begin{equation}
\label{IW83}
\rho(t) = \frac{\pi^{2}}{30} g_{\ast} T^{4}(t)
\end{equation}
where, from (\ref{IW50c}) and (\ref{IW79}),
\begin{equation}
\label{IW84}
\rho(t)=\rho_{0} a(t)^{-3(1+w)} = \rho_{0}\left[\frac{\nu \gamma \sqrt{\lambda_{\gamma} \rho_{0}^{\gamma}}}{2} t \right]^{-2/\gamma}.
\end{equation}
The time-Temperature relationship is therefore
\begin{equation}
\label{IW85}
t = \left[\frac{30}{\pi^{2} g_{\ast}}\right]^{\gamma/2} 
\left[\frac{2}{\nu \gamma \sqrt{\lambda_{\gamma}}}\right] T^{-2 \gamma}.
\end{equation}
Substitution of this relationship into (\ref{IW82}) gives $\dot{R}$ as a function of temperature:
\begin{eqnarray}
\label{IW86}
\dot{R}(T) & = & -3 \lambda_{\gamma}^{3/2} \left[\frac{\pi^{2}g_{\ast}}{30}\right]^{3 \gamma/2}\nu \gamma (4-\nu \gamma) T^{6 \gamma}
\nonumber  \\
& = & -\xi_{\gamma}(w,\beta) T^{6\gamma},
\end{eqnarray}
where
\begin{equation}
\label{IW86a}
\xi_{\gamma}(w,\beta) \equiv 9\lambda_{\gamma}^{3/2} \left[\frac{\pi^{2}g_{\ast}}{30}\right]^{3 \gamma/2} (1+w) \gamma [4-3\gamma (1+w)].
\end{equation}

Hence
\begin{enumerate}
\item
The GB regime: 
\begin{equation}
\label{IW86b}
\dot{R}(T) = -3 \frac{1+\beta}{\beta} \left[\frac{\pi^{2} g_{\ast}}{30}\right](1+w)(1-w) \mu \kappa_{4}^{2}T^{4}.
\end{equation}
\item
The RS regime: 
\begin{equation}
\label{IW86c}
\dot{R}(T) = 36 \left[\frac{\pi^{2} g_{\ast}}{30}\right]^{3}(1+w)(1+3w) \left[\frac{\kappa_{4}^{2}}{6 m_{\sigma}^{4}}\right]^{3/2} T^{12}.
\end{equation}
\item
The GR regime: 
\begin{equation}
\label{IW86d}
\dot{R}(T) = -9 \left[\frac{\pi^{2} g_{\ast}}{30}\right]^{3/2}(1+w)(1+3w) \left[\frac{\kappa_{4}^{2}}{3}\right]^{3/2} T^{6}.
\end{equation}

\end{enumerate}

The baryon asymmetry factor as a function of temperature is then
\begin{equation}
\label{IW87}
\eta(T)= \left[\frac{15 g_{\mathrm{B}}}{4 \pi^{2} g_{\ast}}\right] \xi_{\gamma}(w,\beta) 
\left[\frac{T^{6\gamma-1}}{M_{\ast}^{2}}\right].
\end{equation}
That is
\begin{enumerate}
\item
The GB regime: 
\begin{equation}
\label{IW87a}
\eta(T) = \frac{3(1+\beta)}{4 \beta} (1+w)(1-w)\left(\frac{\mu}{M_{\mathrm{Pl}}}\right) 
\left(\frac{T^{3}}{M_{\mathrm{Pl}}M_{\ast}^{2}}\right).
\end{equation}
\item
The RS regime: 
\begin{eqnarray}
\label{IW87b}
\eta(T)& = & -\frac{3}{2\sqrt{6}} \left[\frac{\pi^{2} g_{\ast}}{30}\right]^{2}(1+w)(1+3w)
\nonumber  \\ 
& & \times \left(\frac{M_{\mathrm{Pl}}}{m_{\sigma}}\right)^{6} \left(\frac{T^{11}}{M_{\ast}^{2}M_{\mathrm{Pl}}^{9}}\right).
\end{eqnarray}
\item
The GR regime: 
\begin{equation}
\label{IW87c}
\eta(T) = \sqrt{3} \left[\frac{\pi^{2} g_{\ast}}{30}\right]^{1/2}(1+w)(1-3w) 
\left(\frac{T^{5}}{M_{\ast}^{2}M_{\mathrm{Pl}}^{3}}\right).
\end{equation}

\end{enumerate}

Note that, for the RS regime, $\dot{R} >0$, indicative of the unconventional cosmology of RS braneworlds~\cite{Langlois2003,BDL2000}.
Hence, for successful baryogenesis, the RS regime must be avoided, either by ensuring the decoupling temperature occurs during 
the GB regime, or by suppressing the RS regime so that the Universe makes a direct transition from the GB regime to the GR regime.
This requires $m_{\alpha} = m_{\sigma}$, that is $\mathcal{R}_{m}=1$, or $\beta = 0.1509$.

Defining the transition between the GB and RS regimes by $\rho(T_{\mathrm{Tr}})=m_{\alpha}^{4}$ then
\begin{equation}
\label{IW87f}
T_{\mathrm{Tr}} =\left[ \frac{60 M_{\mathrm{Pl}}^{2}}{\pi^{2} g_{\ast}}\right]^{1/4} \sqrt{\mu}
 \left[\frac{2(1-\beta)^{3}}{\beta (1+\beta)^{2}}\right]^{1/8}.
\end{equation}

The decoupling temperature $T_{\mathrm{D}}$ required to produce the observed asymmetry $\eta_{\mathrm{obs}}=9 \times 10^{-11}$
is, from (\ref{IW87}),
\begin{equation}
\label{IW87d}
T_{\mathrm{D}} = \left[\frac{4 \pi^{2} g_{\ast}}{15 g_{\mathrm{B}}}
\frac{M_{\ast}^{2}\eta_{\mathrm{obs}}}{\xi_{\gamma}(w,\beta)}\right]^{1/(6 \gamma-1)}.
\end{equation}

\section{Gravitational baryogenesis in the GB regime}

The baryon asymmetry factor (\ref{IW87a}) depends upon the physical constants $M_{\ast}$ and $\mu$.  The cutoff scale $M_{\ast}$ is 
usually taken as $M_{\ast}=M_{\mathrm{Pl}}$, and this value will be assumed in the following calculations.

Writing the baryon asymmetry factor (\ref{IW87a}) as
\begin{equation}
\label{IW201}
\eta = \frac{3 \mu}{4} (1+w)(1-w)\chi(\beta)\, \Theta (T),
\end{equation}
where
\begin{equation}
\label{IW202}
\chi(\beta) \equiv \frac{(1+\beta)}{\beta} 
\end{equation}
and
\begin{equation}
\label{IW203}
\Theta (T) \equiv \left(\frac{T^{3}}{M_{\mathrm{Pl}}^{2}M_{\ast}^{2}}\right),
\end{equation}
then, for a decoupling temperature of the order of the inflationary scale $T=T_{\mathrm{D}} = M_{\mathrm{I}}$, where $M_{\mathrm{I}}$
is constrained by the upper bound on tensor mode fluctuations to be $\sim 2 \times 10^{16}$ GeV, 
the factor $\Theta$ has the value $2.276 \times 10^{-25}$ for $M_{\ast}=M_{\mathrm{Pl}}$.

Several studies provide estimates of GB parameters. ~\cite{DLMS2004} note that the limit on deviations from Newton's 
laws require $l \lesssim 0.1$ mm. Since 1 m $= 5.06776 \times 10^{15}$ GeV$^{-1}$, then $l \lesssim 5.068 \times 10^{11}$ GeV$^{-1}$.
From (\ref{IW58}), an estimate of the brane tension is
\begin{eqnarray}
\label{IW107}
\sigma  & =& \frac{2 \mu}{\kappa_{5}^{2}}\left(3-\frac{4 \alpha}{l^{2}}\right) \approx \frac{6 \mu}{\kappa_{5}^{2}}
= \frac{6 \mu^{2}}{\kappa_{4}^{2}} = \frac{6}{l^{2}\kappa_{4}^{2}}
\nonumber  \\
& \geq & 1.38 \times 10^{14}\,\mathrm{GeV}^{4}.
\end{eqnarray}
Defining
\begin{equation}
\label{IW106a}
M_{5}^{3}\equiv \frac{8 \pi}{\kappa_{5}^{2}}, \quad M_{4}^{2} \equiv \frac{8 \pi}{\kappa_{4}^{2}},
\end{equation}
where, from, (\ref{IW70}), 
\begin{equation}
\label{IW106}
\kappa_{4}^{2} = \frac{\mu}{1+4 \alpha/l^{2}} \kappa_{5}^{2} \simeq \mu \kappa_{5}^{2}
\end{equation}
since $\alpha \ll l^{2}$, then $M_{5}^{3} \approx \mu M_{4}^{2} \approx M_{4}^{2}/l$, and therefore 
$M_{5} \geq 6.65 \times 10^{8}$ GeV.

Another requirement is that, for GB to be a correction to RS gravity, $m_{\alpha}$ must be greater than the RS energy scale $m_{\sigma}$, that is $\mathcal{R}_{m}>1$, which requires $\beta <0.1509$.
~\cite{DLMS2004} use $\beta =10^{-3}$ in their study of tensor and scalar perturbations.

Bounds on $\sigma$ and $\alpha$ come from ~\cite{LN2003} in their study of inflation in GB cosmology. They assume (i) the inflaton 
potential $V(\phi)$ dominates the brane tension, that is, $V(\phi) \approx \rho$, and (ii) slow-roll inflation, which is described by 
the parameters $\epsilon =-\dot{H}/H^{2}$, $\eta =V^{\prime\prime}/3H^{2}$, and the spectral index of the scalar spectrum
$n_{s}=1-6\epsilon +2 \eta$. For $H^{2}=\lambda \rho^{2/3}$ and a scale-invariant spectrum ($n_{s}=1$), the 
potential has the exponential form $V =V_{0} e^{\omega \kappa_{4}\phi}$ where $\omega$ determines the self-coupling of the field. 
The values of the brane tension $\sigma$ and GB coupling $\alpha$ that are 
consistent with COBE normalization can then be determined. They find $10^{51} \leq \sigma \leq 10^{58}$ GeV$^{4}$ and 
$10^{-30} \leq \alpha \leq 10^{-24}$ GeV$^{-2}$ for the variable $\chi$ evaluated 70 e-folds before the end of inflation, 
and for $\omega=10,\,100,\,500$. 

\cite{OO2015} have studied the implications of the Planck 2015 measurements on simple inflationary models in GB braneworld
cosmology. They assume $V(\phi)=\rho$, slow-roll inflation, and impose the condition $m_{\alpha}=m_{\sigma}$, that is, 
$\beta=0.1509$, in order to collapse the RS regime.  They find that the predictions for the slow-roll parameters $n_{s}$ and $r$  
for the chaotic inflation potential $V(\phi)=m^{2}\phi^{2}/2$ lie inside the 95\% C.L. contours for
$1.89 \times 10^{12} \leq \mu \leq 3.44 \times 10^{12}$ GeV for 60 e-folds. For a Higgs potential 
$V(\phi)=\lambda(\phi^{2}-v^{2})^{2}$, they find the lower bounds $\mu \geq (5.13,3.41,1.03) \times 10^{12}$ GeV for $v=(20,30,100)$.

\cite{Neupane2014} has studied the implications of Planck 2013 and BICEP2~\cite{BICEP2} data on GB assisted braneworld inflation. This
study uses the GB parameters $\hat{\alpha}=\alpha M_{5}^{2}$ and $\lambda =\Lambda_{5}/(3 M_{5}^{2})$, does not apply
the fine tuning (\ref{IW58}) of the brane tension, and assumes slow-roll conditions. They find $M_{5} \simeq 5.51 \times 10^{16}$ GeV
and, for $V(\phi)=m^{2}\phi^{2}$,  that during inflation where $m \geq H \simeq 1 \times 10^{14}$ GeV,
the constraints $\hat{\alpha} \sim 91-364$ and $\beta \sim (1-4) \times 10^{-4}$. For a small mass inflaton field
$m \simeq 2 \times 10^{13}$ GeV, the bound on $\beta$ is tighter, $\beta \sim (25-130) \times 10^{-4}$. 
Taking $\hat{\alpha} \sim 228$, that is, $\alpha \sim 7.51 \times 10^{-32}$, which is roughly comparable to the 
estimate of ~\cite{LN2003}, then 
$\mu =\sqrt{\beta/(4 \hat{\alpha})} M_{5} \sim (3.3 - 38) \times 10^{-4} M_{5}  \sim (1.8 -21) \times 10^{13}$ GeV.

Using this constraint on $\mu$, the asymmetry is
\begin{equation}
\label{IW204}
\eta >  (3.1 -36) \times 10^{-12} (1+w)(1-w) \chi(\beta). 
\end{equation}
Noting that $\chi(\beta) = 7.627$  for $\beta =0.1509$ and, for $\beta =10^{-n}$, $\chi(\beta) =1+10^{n} \approx 10^{n}$ for $n \geq 3$,
then, in order to obtain $\eta \sim 10^{-10}$, the parameter $\beta$ would have to be $O(10^{-2}-10^{-3})$.

Guided by the above constraints on $\mu$ and $\beta$, the asymmetry (\ref{IW201}) has been calculated for 
$\mu = (10^{12}, 10^{13}, 10^{14})$ GeV, $w=(0,0.1,0.2,0.3,1/3)$ and $\beta =(0.1509,10^{-1},10^{-2},10^{-3},10^{-4})$.
The results are shown in Table \ref{tab:eta}. It is apparent that the observed values (\ref{IW1}) 
$\eta_{\mathrm{obs}} =(8-9) \times 10^{-11}$ can
be obtained for a value of $\beta$ within the range  $(10^{-2} - 10^{-3})$ for $\mu =10^{12}$ GeV, and within 
the range $(10^{-1}- 10^{-2})$ for $\mu =10^{13}$ GeV. All values of $\eta$ for $\mu=10^{14}$ are too high.

\begin{table*}
\caption{\label{tab:eta}Baryon asymmetries $\eta$ for a decoupling temperature $T_{\mathrm{D}}=2 \times 10^{16}$ GeV and 
cut off scale $M_{\ast}=M_{\mathrm{Pl}}$}
\begin{ruledtabular}
\begin{tabular}{lllllll}
$\mu$ & $\beta/w$ & 0.0 & 0.1 & 0.2 & 0.3 & 1/3 \\
\hline
$10^{12}$ & 0.1509 & 1.302(-12) & 1.289(-12) & 1.250(-12) & 1.185(-12) & 1.157(-12) \\
& $10^{-1}$ & 1.877(-12) & 1.859(-12) & 1.802(-12) & 1.708(-12) & 1.669(-12)  \\
& $10^{-2}$ & 1.724(-11) & 1.707(-11) & 1.655(-11) & 1.569(-11) & 1.532(-11)    \\
& $10^{-3}$ & 1.708(-10) & 1.691(-10) & 1.640(-10) & 1.555(-10) & 1.519(-10)    \\
& $10^{-4}$ & 1.707(-9)  & 1.690(-9)  & 1.639(-9)  & 1.553(-9)  & 1.517(-9)  \\
$10^{13}$ & 0.1509 & 1.302(-11) & 1.289(-11) & 1.250(-11) & 1.185(-11) & 1.157(-11) \\
& $10^{-1}$ & 1.877(-11) & 1.859(-11) & 1.802(-11) & 1.708(-11) & 1.669(-11)  \\
& $10^{-2}$ & 1.724(-10) & 1.707(-10) & 1.655(-10) & 1.569(-10) & 1.532(-10)  \\
& $10^{-3}$ & 1.708(-9)  & 1.691(-9)  & 1.640(-9)  & 1.555(-9)  & 1.519(-9)  \\
& $10^{-4}$ & 1.707(-8)  & 1.690(-8)  & 1.639(-8)  & 1.553(-8)  & 1.517(-8)  \\
$10^{14}$ & 0.1509 & 1.302(-10) & 1.289(-10) & 1.250(-10) & 1.185(-10) & 1.157(-10)  \\
& $10^{-1}$ & 1.877(-10) & 1.859(-10) & 1.802(-10) & 1.708(-10) & 1.669(-10)  \\
& $10^{-2}$ & 1.724(-9)  & 1.707(-9)  & 1.655(-9)  & 1.569(-9)  & 1.532(-9  \\
& $10^{-3}$ & 1.708(-8)  & 1.691(-8)  & 1.640(-8)  & 1.555(-8)  & 1.519(-8)   \\
& $10^{-4}$ & 1.707(-7)  & 1.690(-7)  & 1.639(-7)  & 1.553(-7)  & 1.517(-7)  \\
\end{tabular}
\end{ruledtabular}
\end{table*}

The decoupling temperatures $T_{\mathrm{D}}$ required in the GB regime to produce the observed asymmetry $\eta_{\mathrm{obs}}$
can be calculated from (\ref{IW201}), which gives
\begin{equation}
\label{IW205}
T_{\mathrm{D}} = \left[\frac{4 \eta_{\mathrm{obs}} M_{\mathrm{Pl}}^{2}m_{\ast}^{2}}{3(1+w)(1-w) \chi (\beta)}\right]^{1/3}.
\end{equation}
Values of $T_{\mathrm{D}}$ for the same range of parameters used above are shown in Table \ref{tab:TD}. Also shown are the transition 
temperatures $T_{\mathrm{Tr}}$ (GeV) calculated from (\ref{IW87f}) representing the end of the GB regime.
The requirement $T_{\mathrm{Tr}} \leq T_{\mathrm{D}} \leq M_{\mathrm{I}}$, where $M_{\mathrm{I}}\sim 2 \times 10^{16}$ GeV,
is satisfied for $\mu ,\beta$ values $(10^{12},10^{-4})$, $(10^{13}, 10^{-3})$ and $(10^{14}, 10^{-2})$.

\begin{table*}
\caption{\label{tab:TD}Decoupling temperatures $T_{\mathrm{D}}$ (GeV)  required to produce the observed baryon asymmetry 
$\eta_{\mathrm{obs}} =9 \times 10^{-11}$ for a cut off scale $M_{\ast}=M_{\mathrm{Pl}}$. Also shown are the transition 
temperatures $T_{\mathrm{Tr}}$ (GeV) representing the end of the GB regime}
\begin{ruledtabular}
\begin{tabular}{llllllll}
$\mu$ & $\beta/w$ & 0.0 & 0.1 & 0.2 & 0.3 & 1/3 & $T_{\mathrm{Tr}}$ \\
\hline
$10^{12}$ & 0.1509 & 1.769(17) & 1.774(17) & 1.793(17) & 1.825(17) & 1.839(17) & 9.561(14)\\
& $10^{-1}$ & 1.565(17) & 1.571(17) & 1.587(17) & 1.615(17) & 1.628(17) & 1.040(15) \\
& $10^{-2}$ & 7.475(16) & 7.500(16) & 7.578(16) & 7.714(16) & 7.774(16) & 1.469(15)   \\
& $10^{-3}$ & 3.480(16) & 3.492(16) & 3.528(16) & 3.591(16) & 3.619(16) & 1.970(15)   \\
& $10^{-4}$ & 1.616(16) & 1.621(16) & 1.638(16) & 1.667(16) & 1.680(16) & 2.629(15) \\
$10^{13}$ & 0.1509 & 8.209(16) & 8.236(16) & 8.321(16) & 8.471(16) & 8.537(16) & 3.024(15) \\
& $10^{-1}$ & 7.265(16) & 7.290(16) & 7.365(16) & 7.497(16) & 7.556(16) & 3.290(15) \\
& $10^{-2}$ & 3.470(16) & 3.481(16) & 3.517(16) & 3.580(16) & 3.609(16) & 4.645(15) \\
& $10^{-3}$ & 1.615(16) & 1.621(16) & 1.637(16) & 1.667(16) & 1.680(16) & 6.230(15) \\
& $10^{-4}$ & 7.500(15) & 7.525(15) & 7.602(15) & 7.739(15) & 7.800(15) & 8.312(15) \\
$10^{14}$ & 0.1509 & 3.810(16) & 3.823(16) & 3.862(16) & 3.932(16) & 3.963(16) & 9.561(15) \\
& $10^{-1}$ & 3.372(16) & 3.384(16) & 3.419(16) & 3.480(16) & 3.507(16) & 1.040(16) \\
& $10^{-2}$ & 1.610(16) & 1.616(16) & 1.633(16) & 1.662(16) & 1.675(16) & 1.469(16) \\
& $10^{-3}$ & 7.497(15) & 7.523(15) & 7.600(15) & 7.737(15) & 7.798(15) & 1.970(16) \\
& $10^{-4}$ & 3.481(15) & 3.493(15) & 3.529(15) & 3.592(15) & 3.620(15) & 2.629(16) \\
\end{tabular}
\end{ruledtabular}
\end{table*}

The Universe is required to remain in thermal equilibrium during gravitational baryogenesis. Using 
\begin{equation}
\label{IW108}
H(t) = \frac{1}{(1+w)t}
\end{equation}
and, from (\ref{IW85}),
\begin{equation}
\label{IW109}
t = \left[\frac{30}{\pi^{2} g_{\ast}}\right]^{1/3} 
\left[\frac{1}{(1+w) \sqrt{\lambda}}\right] T^{-4/3},
\end{equation}
then
\begin{equation}
\label{IW110}
H(T)= \left[\frac{\pi^{2}g_{\ast} (1+\beta)}{120 \beta}\right]^{1/3} (\mu \kappa_{4}^{2})^{1/3} \,T^{4/3}.
\end{equation}
Requiring $\Gamma_{\mathrm{B}} = T^{2n+1}/M_{\mathrm{B}}^{2n}> H$ then $M_{\mathrm{B}}$ must satisfy
\begin{equation}
\label{IW111}
M_{\mathrm{B}} < \left[ \frac{\pi^{2}g_{\ast} (1+\beta)}{120 \beta}\right]^{6n} (\mu \kappa_{4}^{2})^{-1/6n}\,
T^{1-1/6n}.
\end{equation}
For a dimension-six operator $(n=2)$, this gives
\begin{equation}
\label{IW112}
M_{\mathrm{B}} < \left[ \frac{\pi^{2}g_{\ast} (1+\beta)}{120 \beta}\right]^{12} (\mu \kappa_{4}^{2})^{-1/12}\,
T^{11/12}.
\end{equation}
For $\mu \geq 10^{12}$ GeV, $(\mu \kappa_{4}^{2})^{-1/12} \leq 1.16 \times 10^{2}$ GeV$^{1/12}$. 
Choosing $T=T_{\mathrm{D}} =2 \times 10^{16}$ GeV, then $T^{11/12}=8.76 \times 10^{14}$ GeV$^{11/12}$. Hence, 
$M_{\mathrm{B}} < 2.94 \times 10^{28} [(1+\beta)/\beta]^{1/12}$ GeV.  For $\beta=10^{-n}$ this gives 
$M_{\mathrm{B}} < 2.94 \times 10^{28+n/12}$ GeV. This thermalization requirement is clearly satisfied by the expected range
$M_{\mathrm{B}} \sim O(10^{14} - 10^{16})$ GeV for $n>0$.

\section{Summary and discussion}

The feasibility of gravitational baryogenesis in a Gauss-Bonnet (GB) modified Randall-Sundrum (RS) type II braneworld cosmology
has been investigated. The modified expansion rate of the Universe in this scenario evolves through three distinct expansion regimes. 
At early times, during the GB regime, the expansion rate of the universe $H \sim \rho^{1/3}$ falls more slowly than the standard expansion law $H\sim \rho^{1/2}$ of General Relativity (GR). Later, the universe evolves into a RS type era with an enhanced expansion $H \sim \rho$, before finally reducing to the standard expansion law in the low energy limit. 
The duration of the RS regime is determined by the magnitude of the parameter $\beta$ and is completely 
suppressed if $\beta =0.1509$.

The gravitational baryogenesis is calculated assuming the gravitational interaction (\ref{IW2}) does not affect the cosmological 
background dynamics, and that the $B$ number-violating interactions are more rapid than the expansion rate of the Universe, thus 
ensuring the Universe remains in thermal equilibrium. Gravitational baryogenesis depends on $\dot{R}$ and requires $\dot{R} <0$. 
This condition is satisfied for the GB and GR regimes. However, $\dot{R}>0$ for the RS regime and successful baryogenesis 
requires either that the decoupling temperature occurs during the GB regime or that the Universe makes a direct transition 
from the GB regime to the GR regime.

The change of sign of $\dot{R}$ for the RS regime is not commented on by Bento \textit{et al}~\cite{BFS2005} in their 
treatment of gravitational baryogenesis in Gauss-Bonnet braneworld cosmology for the restricted case of a radiation dominated Universe.

The GB baryon asymmetry $\eta(T)$ (\ref{IW201}) for a given decoupling temperature $T=T_{\mathrm{D}}$  
and cutoff scale  $M_{\mathrm{\ast}}$ is proportional to $\mu (1+w)(1-w)(1+1/\beta)$,  
where $\mu$ is the energy scale $\mu \equiv 1/l$ associated with the bulk curvature scale $l$, 
$\beta \equiv 4 \alpha \mu^{2}$, where $\alpha$ is the GB coupling, and $w$ is the equation of state parameter $p=w \rho$.
Studies~\cite{OO2015,Neupane2014} of inflation in GB-brane-world cosmology place constraints on the allowed 
ranges of the parameters $\mu$ and $\beta$. Guided by these, the asymmetry (\ref{IW201}) has been calculated for 
$\mu = (10^{12}, 10^{13}, 10^{14})$ GeV, $w=(0,0.1,0.2,0.3,1/3)$ and $\beta =(0.1509,10^{-1},10^{-2},10^{-3},10^{-4})$,
assuming $T_{\mathrm{D}} =2 \times 10^{16}$ GeV and $M_{\mathrm{\ast}}=M_{\mathrm{Pl}}$.
It is apparent that the observed values (\ref{IW1}) $\eta_{\mathrm{obs}} =(8-9) \times 10^{-11}$ can
be obtained for a value of $\beta$ within the range  $(10^{-2} - 10^{-3})$ for $\mu =10^{12}$ GeV, and within the 
range $(10^{-1}- 10^{-2})$ for $\mu =10^{13}$ GeV. All values of $\eta$ for $\mu=10^{14}$ are too high.

The decoupling temperatures $T_{\mathrm{D}}$ required in the GB regime to produce the observed asymmetry have been calculated 
for the same range of parameters. The requirement $T_{\mathrm{Tr}} \leq T_{\mathrm{D}} \leq M_{\mathrm{I}}$, where the
transition temperature $T_{\mathrm{Tr}}$ represents the end of the GB regime,
is satisfied for $\mu ,\beta$ values $(10^{12},10^{-4})$, $(10^{13}, 10^{-3})$ and $(10^{14}, 10^{-2})$.

Gravitational baryogenesis assumes the Universe is in thermal equilibrium. If the $B$ number-violating interactions 
are generated by a dimension-six operator with an associated mass scale $M_{\mathrm{B}}$, this thermalization requirement 
becomes (\ref{IW112}). The allowed parameter ranges for $\mu$ and $\beta$ found in the present study easily satisfy this 
requirement for $M_{\mathrm{B}}$ in the range $M_{\mathrm{B}} \sim O(10^{14} - 10^{16})$ GeV that is needed to satisfy constraints 
from GUT or Planck-suppressed interactions.

The modified Friedmann equation (\ref{IW68}) for the RS/GR regime has the same form as that (\ref{IW89}) for Loop Quantum Cosmology, except that the $\rho^{2}$ term has opposite signs in the two cosmologies. This has the consequence~\cite{OO2016a} that $\dot{R}$ is always negative, as desired for successful generation of a baryon asymmetry, in Loop Quantum Cosmology, but changes sign in GB braneworld cosmology, being positive for the RS regime. 

In contrast to GR where the baryon asymmetry factor is zero for a radiation dominated Universe ($w=1/3$), it is non-zero in GB braneworld cosmology.

The present investigation has shown that gravitational baryogenesis in a Gauss-Bonnet modified Randall-Sundrum type II braneworld
cosmology can generate a baryon asymmetry compatible with the observed value, for ranges of the free parameters which satisfy the 
constraints imposed by CMB measurements, for decoupling temperatures $ \sim 10^{16}$ GeV. 

%\end{acknowledgments}

\end{document}